\documentstyle[]{article}

\font\grb=eurb10

\def\bphi{\hbox{\grb\char'047}\,}
\def\bpsi{\hbox{\grb\char'040}\,}
\def\bchi{\hbox{\grb\char'037}\,}

\def\e{\hbox{\bf e}}
\def\f{\hbox{\bf f}}
\def\g{\hbox{\bf g}}
\def\k{\hbox{\bf k}}
\def\l{\hbox{\bf l}}
\def\m{\hbox{\bf m}}

\def\p{\hbox{\bf p}}

\def\u{\hbox{\bf u}}
\def\v{\hbox{\bf v}}
\def\G{\hbox{\bf G}}
\def\M{\hbox{\bf M}}
\def\N{\hbox{\bf N}}
\def\R{\hbox{\bf R}}
\def\T{\hbox{\bf T}}
\def\U{\hbox{\bf U}}
\def\V{\hbox{\bf V}}
\def\W{\hbox{\bf W}}
\def\bigpsi{{\bf \Psi}}
\def\bigomega{{\bf \Omega}}
\def\vpint{\mathop{\int\hskip-0.8pc\mbox{\it /}}}

\title{New integral equation form of integrable
reductions of Einstein equations}
\author{G. A. Alekseev\\
\normalsize \em
Steklov Mathematical Institute of the Russian Ac. Sci.,\\
\normalsize \em
Gubkina 8, Moscow 117966, GSP-1, Moscow, Russia,\\
\normalsize \em
e-mail: G.A.Alekseev@mi.ras.ru}

\date{}

\begin{document}

\renewcommand{\thepage}{}
\begin{titlepage}
\maketitle
\begin{abstract}
A new development of the ``monodromy transform'' method for
analysis of hyperbolic as well as elliptic integrable reductions
of Einstein equations is presented. Compatibility conditions for
some alternative representations of the fundamental solutions of
associated linear systems with spectral parameter in terms of a
pair of dressing (``scattering'') matrices give rise to a new
set of linear (quasi-Fredholm) integral equations equivalent to
the symmetry reduced Einstein equations.  Unlike previously
derived singular integral equations constructed with the use of
conserved (nonevolving) monodromy data on the spectral plane for
the fundamental solutions of associated linear systems, the
scalar kernels of the new equations include another kind of
functional parameters -- the evolving (``dynamical'') monodromy
data for the scattering matrices. For hyperbolic reductions, in
the context of characteristic initial value problem these data
are determined completely by the characteristic initial data for
the fields. In terms of solutions of the new integral equations
the field components are expressed in quadratures.
\end{abstract}
\vfill
\end{titlepage}
\renewcommand{\thepage}{\arabic{page}}


\section{Introduction}

In General Relativity in a number of physically significant cases
the dynamical part of Einstein equations, being restricted to
space-time geometries admitting two-dimensional Abelian isometry
group\footnote{The most known and elegant form of the reduced
equations is written in terms of complex potentials --- the
Ernst potentials. These are the vacuum Ernst equation for one
complex potential \cite{Ernst:1968a} or electrovacuum Ernst
equations, i.e.  a coupled system of two similar equations for
two complex Ernst potentials \cite{Ernst:1968b} which have been
derived by F.J.Ernst originally for stationary axisymmetric
fields. For the hyperbolic case these equations have very
similar forms \cite{Hauser-Ernst:1989}. A generalized form of
these equations arises, for example, in the presence of Weyl
spinor field \cite{Alekseev:1983}.}, reduces to the nonlinear
integrable systems.  Among these are the Einstein equations for
vacuum gravitational fields
\cite{Maison:1978}--\cite{Hauser-Ernst:1979}, the Einstein
equations for space-times with a stiff matter fluid
\cite{Belinskii:1979}, the electrovacuum Einstein - Maxwell field
equations
\cite{Kinnersley:1977,
Kinnersley-Chitre:1977,Hauser-Ernst:1979,Alekseev:1980}, the
Einstein - Maxwell - Weyl equations for gravitational,
electromagnetic and massless spinor fields \cite{Alekseev:1983},
as well as some string theory induced gravity models, e.g., the
Einstein - Maxwell equations with axion and dilaton fields
\cite{Bakas:1994} -- \cite{Alekseev:2000a}.
Accordingly to the type of the contemplated two-dimensional
space-time symmetry (determined by the signature of the metric
on the orbit space) the reduced equations can be either of the
hyperbolic or of the elliptic types. By now a theory of these
equations has been developed and discussed in many aspects (for
details and references see, for example, \cite{Alekseev:2000a,
Ernst:1999}).

The so called monodromy transform
\cite{Alekseev:2000a,Alekseev:1985,Alekseev:1987} provides some
general and fairly simple base for a description
in a unified manner of all mentioned above integrable reductions
of Einstein equations.  Similarly to the well known inverse
scattering method (the ``scattering transform''), this approach
begins with a representation of the  dynamical part of
reduced Einstein equations as the integrability conditions of
an overdetermined linear system of a special structure
containing a spectral parameter. The analysis of the constructed
linear systems showed a common important property of the
evolution of fields described by all mentioned above integrable
reductions of Einstein equations.  This property is a
conservation of the monodromy structure on the spectral plane of
the normalized fundamental solutions of associated linear
systems.  This monodromy structure is defined here as a set of
linear transformations which relate this fundamental solution
with its analytical continuations along the closed paths
surrounding each of the singular points of this solution on the
spectral plane. The matrices of these linear transformations
possess some special algebraic structure with a small set of
independent components -- the functions of the spectral
parameter which constitute what we call here as monodromy
data.\footnote{For example, for any vacuum gravitational field with
the supposed space-time symmetry the monodromy data consist of
two functions of the spectral parameter, while for elctrovacuum
fields we have four such functions.} A remarkable properties of
the defined monodromy data are that they (i) are functions of
the spectral parameter only, (ii) exist for any analytical
local solution of reduced Einstein equations and (iii)
characterize uniquely every analytical local solution of the
field  equations. All these means, that these monodromy data can
be considered as a new set of the field variables for which the
field equations become trivial (being the conserved quantities,
these monodromy data do not possess any evolution) or as new
``coordinates'' (instead of usual field variables) in the
infinite dimensional spaces of local solutions of the integrable
reductions of Einstein equations. Just using of such
``coordinate transformation'' explains why this approach was
called as the monodromy transform.

In this way the ``direct'' and ``inverse'' problems of the
monodromy transform suggest themselves naturally. For solution
of the first of them, i.e. for calculation of the monodromy data
for a given solution of reduced Einstein equations, we have to
find a fundamental solution of the associated linear system
with the coefficients corresponding to given field components
and to determine than its monodromy data on the spectral plane.
The solution of the inverse problem, i.e. the construction of a
solution of Einstein equations for given monodromy data
functions, reduces to the solution of a special integral
equation form of the field equations derived in
\cite{Alekseev:1985,Alekseev:1987}.  These are the linear
singular integral equations whose scalar kernels and right hand
sides are expressed in terms of the mentioned above nonevolving
monodromy data. (Equivalent regularizations of these equations
have been derived in
\cite{Hauser-Ernst:2001,Alekseev:2000a,Alekseev:2000b}.) These
integral equations always possess a unique solution for any
given monodromy data functions holomorphic in some local regions
of the spectral plane.  The corresponding local solution of
Einstein equations can be calculated in quadratures whose
integrands are expressed in terms of the solution of the
mentioned linear integral equations.

This approach suggests different applications. The first of them is a
direct construction of local solutions of Einstein equations. In
this case the actual problem is to chose some monodromy data
which allow to solve explicitly the integral equations
with the corresponding kernels and then to calculate the
corresponding quadratures for the field components. Fortunately, for
very large classes of the monodromy data all these can be
realized, and infinite hierarchies of families of exact
solutions with any finitely large number of free
parameters can be calculated in elementary functions. The
examples of such hierarchies are the solutions with arbitrary
analytically adjusted rational monodromy data
\cite{Alekseev:1987,Alekseev:1992,Alekseev-Garcia:1996} and the
solutions for colliding plane waves and
inhomogeneous cosmological models with analytically not adjusted
but also rational monodromy data
\cite{Alekseev-Griffiths:2000}. Each of these hierarchies
extends considerably the hierarchies of vacuum
\cite{Belinskii-Zakharov:1978} and electrovacuum
\cite{Alekseev:1980} multisoliton solutions provided
the Minkowski space-time is chosen as the background
for solitons.

Another application of the same approach is a formulation of general
schemes for solution of some initial or boundary value problems for
the gravitational and some matter fields with two--dimensional
space-time symmetries. The basic idea is that any given initial
data for a characteristic initial value problem or a Cauchy
problem for integrable hyperbolic reductions of Einstein
equations as well as a boundary data for some boundary
problems for an elliptic (e.g., stationary axisymmetric)
integrable reduction of Einstein equations allow to calculate
the corresponding monodromy data. The conserved (i.e.
nonevolving) character of these data allows to identify them
with the monodromy data for the sought-for solution of the
initial or boundary value problem under consideration.

In the present paper a derivation of a new form of integral
equations equivalent to each of the mentioned above
integrable reductions of Einstein equations is presented. For
this we introduce a specific alternative
representations of the fundamental solutions of the associated
linear systems in terms of pairs of ``scattering''  matrices
dressing some partial values of these fundamental solutions
(``in-states''), each depending on one of two coordinates on the
orbit space and on the spectral parameter.  These scattering
matrices were found to possess a specific analytical structure
on the spectral plane, where each of them is characterized by
two (algebraic in the absence of a Weyl spinor field)
branchpoints and finite jump on the cut which joins these
points. (It is useful to recall here that the fundamental
solution of any of the associated linear systems under
consideration in general possess four such branchpoints which
are joined in our construction by two nonintersecting local
cuts.) The consistency conditions for these
alternative representations of solutions of associated linear
systems give rise to linear (quasi-Fredholm) integral equations
interrelating some fragments of the algebraic structures of the
dressing matrices on the cuts. The scalar kernels of these
equations and their right hand sides include functional
parameters which characterize the monodromy properties of the
dressing (scattering) matrices mentioned above.  Unlike the
nonevolving monodromy data for fundamental solutions of
associated linear systems, these (``dynamical'') monodromy data
evolve and their evolution  is determined by a coordinate
dependence of the ``in-states''. In terms of solutions of these
new integral equations all of the field components and the Ernst
potentials, characterizing the solutions, are determined in
quadratures.  For hyperbolic reductions, in the context of
characteristic initial value problem these
``in-states'' can be identified easily with the characteristic
initial data for fundamental solution of associated linear
system which are determined completely by the characteristic
initial data for the fields.

For simplicity, we restrict all considerations following below by
hyperbolic as well as elliptic reductions of vacuum Einstein
equations and electrovacuum Einstein - Maxwell equations only,
because there are no any principal difficulties for the
realization of the constructions suggested below for all other
mentioned above integrable reductions of Einstein equations.

\section{Reduction of Einstein equations}

In this section we describe basic
definitions used below and present in unified notations
the dynamical parts of reduced vacuum Einstein
equations and electrovacuum Einstein - Maxwell equations in
one of their most compact forms --- in the form of the Ernst
equations.

\subsection{Metric and electromagnetic potential}

The components of a 4-dimensional space-time metric and a 1-form
$\underline{\Phi}$ of complex electromagnetic potential for a
self-dual Maxwell 2-form in a space-time admitting
2-dimensional Abelian isometry group can be considered in the
form
\begin{equation}\label{Components} ds^2=g_{\mu\nu}dx^\mu
dx^\nu+g_{ab}dx^a dx^b,\qquad \underline{\Phi}=\Phi_a dx^a
\end{equation}
where $\mu,\nu,\ldots=1,2$; $a,b,\ldots=3,4$; $g_{\mu\nu}$,
$g_{ab}$ and $\Phi_a$ depend on the coordinates $x^1$ and $x^2$
only.  We consider metric $g_{\mu\nu}$ locally in a conformally
flat form and parametrize the nonzero field components in
(\ref{Components}) by scalar functions
\begin{equation}\label{Parametrization}
g_{\mu\nu}=f\left(\begin{array}{cc}
\epsilon_1&0\\0&\epsilon_2\end{array}\right),\quad
g_{ab}=\epsilon_0\left(\begin{array}{ll} H& H\Omega\\ H\Omega&
H\Omega^2+\epsilon\alpha^2/H\end{array}\right),\quad\Phi_a=(
\begin{array}{ll} \Phi,&\widetilde{\Phi}\end{array})
\end{equation}
where $f\ge 0$, $\alpha\ge 0$, $H\ge 0$;
$\epsilon,\epsilon_0,\epsilon_1,\epsilon_2=\pm 1$ are the sign
symbols whose choice should provide the Lorentz signature of the
metric (\ref{Components}). This condition is equivalent to
the relation $\epsilon_1\epsilon_2=-\epsilon$. The sign
symbol $\epsilon$ determines both, the signature of metric $g_{ab}$
on the 2-dimensional orbits of the space-time isometry group
and the signature of the conformally flat metric $g_{\mu\nu}$ on
the two-dimensional orbit space of this group. We shall refer to
the cases $\epsilon=1$ and $\epsilon=-1$ as the hyperbolic and
elliptic ones respectively.

The function $\alpha(x^1,x^2)$ in (\ref{Parametrization})
satisfies the identity $\det\Vert g_{ab}\Vert\equiv
\epsilon\alpha^2$, which means that this function characterizes
a measure of area on the orbits of the isometry group. For all
known integrable reductions of Einstein equations considered
here $\alpha(x^1,x^2)$ is a harmonic function, and this permits
to define a function $\beta$ as its harmonic conjugation as
follows
$$\begin{array}{l}
\eta^{\mu\nu}\partial_\mu\partial_\nu\alpha=0,\\[1ex]
\partial_\mu\beta=-\epsilon\varepsilon_\mu{}^\nu\partial_\nu\alpha,
\end{array}\qquad
\eta^{\mu\nu}=\left(\begin{array}{cc}\epsilon_1&0\\0&\epsilon_2
\end{array}\right)\qquad
\varepsilon_\mu{}^\nu=\left(\begin{array}{cc}
0&\epsilon_1\\-\epsilon_2&0
\end{array}\right)$$
The functions $(\alpha,\beta)$ constitute a set of
``geometrically defined'' coordinates which can be used instead
of unspecified coordinates $(x^1,x^2)$.\footnote{In the
stationary axisymmetric case these geometrically defined
coordinates are known as cylindrical Weyl coordinates
$(\rho,z)$.} However, we use instead of coordinates $\alpha$ and
$\beta$ their linear combinations $(\xi,\eta)$:
$$\left\{\begin{array}{l}
\xi=\beta+j\alpha\\ \eta=\beta-j\alpha \end{array}\right.
\quad\mbox{where}\quad j=\left\{\begin{array}{lcl} 1&\mbox{for}&
\epsilon=1\\ i&\mbox{for}&\epsilon=-1 \end{array}\right.$$
which are two real null cone coordinates in the hyperbolic case
($\epsilon=1$) or complex conjugated to each other coordinates
in the elliptic case ($\epsilon=-1$).

The existing gauge freedom permits without any loss of generality to
impose on the metric and electromagnetic functions the normalization
conditions
\begin{equation}\label{MetricNorm}
H(\xi_0,\eta_0)=1,\quad
\Omega(\xi_0,\eta_0)=0,\quad \Phi(\xi_0,\eta_0)=0
\end{equation}
where $(\xi_0,\eta_0)$ are the coordinates of a
chosen ``reference'' point $P_0$.

\subsection{The Ernst equations}

The dynamical part of the space-time symmetry reduced
electrovacuum Einstein - Maxwell field equations
for  hyperbolic as well as for elliptic cases can be expressed
in the form of the Ernst equations. In our
notations these equations and the linear
equation for $\alpha$, can be written as
\begin{equation}\label{ErnstEquations}
\left\{\begin{array}{l}
(\mbox{Re }{\cal E}+\overline{\Phi}\Phi) \eta^{\mu\nu}(\partial_\mu+
\alpha^{-1}\partial_\mu\alpha) \partial_\nu {\cal
E}-\eta^{\mu\nu} (\partial_\mu{\cal E}
+2\overline{\Phi}\partial_\mu\Phi)\partial_\nu{\cal E}=0\\[1ex]
(\mbox{Re }{\cal E}+\overline{\Phi}\Phi)\eta^{\mu\nu}(\partial_\mu+
\alpha^{-1}\partial_\mu\alpha)
\partial_\nu\Phi-\eta^{\mu\nu} (\partial_\mu{\cal E}
+2\overline{\Phi}\partial_\mu\Phi)\partial_\nu\Phi=0\\[1ex]
\eta^{\mu\nu}\partial_\mu\partial_\nu\alpha=0
\end{array}\right.
\end{equation}
where ${\cal E}(x^1,x^2)$ and $\Phi(x^1,x^2)$ are complex Ernst
potentials, and for vacuum $\Phi\equiv 0$.
The Ernst potential ${\cal E}$ is defined by the
expressions \cite{Ernst:1968a,Ernst:1968b}:
$$\mbox{Re\,}{\cal
E}=\epsilon_0 H-\overline{\Phi}\Phi,\qquad
\partial_\mu(\mbox{Im\,}{\cal E})=-\alpha^{-1}
H^2\varepsilon_\mu{}^\nu \partial_\nu\Omega+i(\overline{\Phi}
\partial_\mu\Phi- \Phi \partial_\mu\overline{\Phi}),$$
while the electromagnetic Ernst potential $\Phi$ coincides with
the corresponding component of the self-dual
electromagnetic potential shown in (\ref{Parametrization}).  The
Ernst equations (\ref{ErnstEquations}) are quasilinear equations
of the hyperbolic type for $\epsilon=1$, and they are of the
elliptic type for $\epsilon=-1$. In the coordinates $\xi$,
$\eta$ we put $\alpha=(\xi-\eta)/2 j$
 and $\eta^{\mu\nu}=\left(\begin{array}{cc}0&1\\
1&0\end{array}\right)$ with an appropriate choice of the sign of
the conformal factor $f$.

\section{The monodromy transform}

In this section we recall the basic constructions of the monodromy
transform approach developed in
\cite{Alekseev:1985,Alekseev:1987,Alekseev:2000a} and applicable in
a unified manner to the analysis of all integrable reductions of
Einstein equations mentioned above and, in
particular, to the Ernst equations (\ref{ErnstEquations}) for
vacuum and electrovacuum fields. The basic idea of this approach
is the using of a specially defined functional parameters
(characterizing every local solution) as new ``coordinates'' in
the infinite dimensional space of local solutions of reduced
Einstein equations instead of usual field variables. A
remarkable property of these nonevolving parameters, depending
on the spectral parameter only and being interpreted as the
monodromy data of the fundamental solutions of associated linear
systems, is that the field equations do not impose any
constraints on these parameters and therefore, such
``coordinate transformation'' solves completely the field
equations.  Thus, the solution of reduced Einstein
equations becomes equivalent to solution of the inverse
problem of  such ``coordinate transform'' (called as
``monodromy transform''), and solution of this
problem turns out to be equivalent to solution of some linear
singular integral equations with scalar kernels.

\subsection{Associated linear systems}

Among various gauge equivalent linear
systems with constant or coordinate dependent spectral parameters
(see \cite{Alekseev:2001} for more details) we
chose the Kinnersley--like linear system whose appropriately
normalized fundamental solution seems to possess the most
simple general analytical structure on the spectral plane.  A
complete set of matrix relations, equivalent to the reduced
Einstein - Maxwell equations, can be expressed in terms of the
four unknown matrix functions which are $2\times 2$-matrices for
vacuum gravitational fields or $3\times 3$-matrices for
gravitational and electromagnetic fields (for other integrable
cases see \cite{Alekseev:2000a}):
\begin{equation}\label{UVW}
\U(\xi,\eta),\quad\V(\xi,\eta),\quad\W(\xi,\eta,w),\quad
\bigpsi(\xi,\eta,w)\end{equation}
where $w$ is a spectral parameter and $\xi$, $\eta$ are the defined
above real (the hyperbolic case) or complex conjugated to each other
(the elliptic case) coordinates.

The first group of constraints imposed on the matrix functions
(\ref{UVW}) consists of two systems of
linear differential equations for $\bigpsi$ with
algebraic constraints imposed on their (also unknown) matrix
coefficients
\begin{equation}\label{UV-equations}
\left\{\begin{array}{lclcl}
2i(w-\xi)\partial_\xi\bigpsi= \U(\xi,\eta)\bigpsi\\[1ex]
2i(w-\eta)\partial_\eta\bigpsi= \V(\xi,\eta)\bigpsi
\end{array}\quad
\right\Vert \quad
\begin{array}{lcl}
\mbox{rank}\,\U=1, && \mbox{tr}\,\U=i\\[1ex]
\mbox{rank}\,\V=1, && \mbox{tr}\,\V=i
\end{array}
\end{equation}
The second group of constraints provides the existence for these
linear systems of a common Hermitian matrix integral of a
special structure \begin{equation}\label{W-equations}
\left.\left\{ \begin{array}{l}
\bigpsi^\dagger\>\W\> \bigpsi = \W_0(w)\\[1ex]
\W_0^\dagger(w)=\W_0(w)
\end{array}\quad\right\Vert
\quad
\displaystyle{\partial\W\over \partial w}=4 i\bigomega
\quad\right\Vert\quad\bigomega =
\left(\begin{array}{rrr} 0&1&0\\-1&0&0\\ 0&0&0\end{array}\right)
\end{equation}
where "${}^\dagger$" is a Hermitian conjugation such that
$\bigpsi^\dagger(\xi,\eta,w)\equiv\overline{\bigpsi^T(\xi,\eta,
\overline{w})}$ and $\W_0(w)$ is an arbitrary Hermitian matrix
function of the spectral parameter. In a vacuum case the third rows
and columns of all matrices should be omitted.\footnote{In all
previous author's formulations of these groups of conditions,
there were included also an additional condition for
electrovacuum case, that the lower right element of $\W$ should
be equal to $1$.  However, this condition turns out to be pure
gauge one, and it can be satisfied by an appropriate choice of
the normalization conditions.}

The third group consists of pure gauge conditions
imposed without any loss of generality
on the values of $\bigpsi$ at the chosen reference point $P_0$
and $\W_0(w)$:
\begin{equation}\label{Normalization}
\bigpsi(\xi_0,\eta_0,w)={\bf I},\qquad \W_0(w)=4
i(w-\beta_0)\bigomega
+\mbox{diag}\,(-4\epsilon\epsilon_0\alpha_0^2,-4\epsilon_0,1)
\end{equation}
where $\alpha_0=(\xi_0-\eta_0)/2 j$ and
$\beta_0=(\xi_0+\eta_0)/2$.

\subsection{Field components and potentials}

The conditions (\ref{UVW}) -- (\ref{Normalization}) contain all
information about the specific structures of $\U$, $\V$
and $\W$ which these matrices should possess to be
correctly expressed in terms of the components of metric and
electromagnetic potential, about the Ernst potentials and their
relations to the metric and electromagnetic potential
components, about the Ernst equations and all properties of the
function $\alpha$ described above (see \cite{Alekseev:1987} for
some explanations). So, the relations given below are not some
additional constraints imposed on the introduced auxiliary
functions, but they follow directly from  (\ref{UVW}) --
(\ref{Normalization}).  In turn, the relations, given below in
this subsection, take place in general, for any local solution of
the reduced electrovacuum Einstein - Maxwell equations. In
particular, the matrices  $\U$ and $\V$ always possess the
structures $$\begin{array}{l} \U={\cal
F}_U\cdot\widehat{\U}\cdot{\cal F}^{-1}_U \\[2ex] \V={\cal
F}_V\cdot\widehat{\V}\cdot{\cal F}^{-1}_V\end{array}
\quad\left\Vert\quad
{\cal F}_U=\left(\begin{array}{ccc}
1&0&0\\ p_+&1&0\\ q_+&0&1\end{array}\right),\quad
{\cal F}_V=\left(\begin{array}{ccc}
1&0&0\\ p_-&1&0\\ q_-&0&1\end{array}\right)\right.$$
where the scalar functions $p_\pm\equiv \Omega\mp\displaystyle{i
j\alpha\over \epsilon_0 H}$, \,
$q_\pm=2\overline{\widetilde{\Phi}}-2\overline{\Phi} p_\pm$ and
$$\widehat{\U}=\left(\begin{array}{c} 1\\ 0\\ 0\end{array}\right)
\otimes\left(\begin{array}{lll} i,&-\partial_\xi{\cal
E},&\partial_\xi\Phi\end{array}\right)\qquad
\widehat{\V}=\left(\begin{array}{c} 1\\ 0\\ 0\end{array}\right)
\otimes\left(\begin{array}{lll} i,&-\partial_\eta{\cal
E},&\partial_\eta\Phi\end{array}\right)
$$
Here ${\cal E}$ and $\Phi$ have to be identified with he Ernst
potentials which characterize any local solution of reduced
vacuum Einstein or electrovacuum Einstein - Maxwell equations.
The matrix function $\W$ is linear with respect to the spectral
parameter $w$, and its components are algebraically expressed in
terms of the metric and complex electromagnetic vector potential
components:  $$ \W=4 i(w-\beta)\bigomega+\G,\qquad
\G=\left(\begin{array}{ll} -4 h^{ab}+4\Phi^a\overline{\Phi}{}^b
&-2\Phi^a \\[1ex] -2\overline{\Phi}{}^b & 1\end{array}\right) $$
where $h^{ab}=\epsilon\alpha^2 g^{ab}$, $g^{ab}$ is the matrix
inverse for the metric components  $g_{ab}$, the column
$\Phi^a= (\begin{array}{ll} \widetilde{\Phi},&
-\Phi\end{array})^T$ with the superscript ${}^T$ meaning a
transposition.  Accordingly to
(\ref{MetricNorm}) and (\ref{Normalization}), $\W_0(w)$
coincides with the value of $\W$ at the reference point (the
normalization point) $P_0$, namely $\W_0(w)=\W(\xi_0,\eta_0,w)$.

\subsection{The monodromy structure of $\bigpsi(\xi,\eta,w)$}
For any local solution of the field equations under consideration the
corresponding solution of (\ref{UVW})--(\ref{Normalization}) for
$\bigpsi(\xi,\eta,w)$, considered as a function of $w$
for given $\xi$ and $\eta$ running some local domains near their
initial values $\xi_0$ and $\eta_0$ respectively,
possesses a number of universal analytical properties
on the spectral plane \cite{Alekseev:1987,Alekseev:2000a}. In
particular, the components of $\bigpsi(\xi,\eta,w)$ and
$\bigpsi^{-1}(\xi,\eta,w)$ are holomorphic everywhere on the spectral
plane (including $w=\infty$, where $\bigpsi(\xi,\eta,w=\infty)={\bf
I}$) outside two nonintersecting local cuts $L_+$ and $L_-$
which endpoints are two ordered pairs of the branchpoints
$(w=\xi_0, w=\xi)$ and $(w=\eta_0, w=\eta)$ respectively. In the
hyperbolic case these cuts are represented by two nonoverlapping
segments of the real axis on the $w$-plane, while in the
elliptic case these cuts are located symmetrically to each other
with respect to this axis in the upper and lower half-planes
respectively \cite{Alekseev:2000a}. This description of
analytical properties of $\bigpsi(\xi,\eta,w)$ we conclude
presenting general expressions for the local structure of
$\bigpsi$ and $\bigpsi^{-1}$ near the branchpoints and the cuts
$L_\pm$ which join them \cite{Alekseev:1985,Alekseev:1987}:
\begin{equation}\label{LocalStructure}
\left.\begin{array}{ll}
L_+:&\bigpsi(\xi,\eta,w)=
\lambda_+^{-1}\bpsi_+(\xi,\eta,w)\otimes\k_+(w)+
\M_+(\xi, \eta,w)\\[2ex]
&\bigpsi^{-1}(\xi,\eta,w)=
\lambda_+\l_+(w)\otimes\bphi_+(\xi,\eta,w)+\N_+(\xi,\eta,w)\\[2ex]
L_-:&\bigpsi(\xi,\eta,w)=
\lambda_-^{-1}\bpsi_-(\xi,\eta,w)\otimes\k_-(w)+
\M_-(\xi,\eta,w)\\[2ex]
&\bigpsi^{-1}(\xi,\eta,w)=
\lambda_-\l_-(w)\otimes\bphi_-(\xi,\eta,w)+\N_-(\xi,\eta,w)
\end{array}
\right\vert
\begin{array}{l}
\lambda_+=\sqrt{\displaystyle{w-\xi\over w-\xi_0}}\\[4ex]
\lambda_-=\sqrt{\displaystyle{w-\eta\over w-\eta_0}}
\end{array}
\end{equation}
where $\lambda_+$ and $\lambda_-$ are holomorphic at $w=\infty$,
provided $\lambda_\pm(w=\infty)=1$, and $\lambda_+$ and $\lambda_-$
possess the jumps on $L_+$ and $L_-$ respectively; all fragments of
these local structures of $\bigpsi$ and $\bigpsi^{-1}$, i.e. each of
the row and column vectors $\k_\pm(w)$ and $\l_\pm(w)$, the row
and column vectors $\bpsi_\pm(\xi,\eta,w)$ and
$\bphi_\pm(\xi,\eta,w)$ and the matrices $M_\pm(\xi,\eta,w)$ and
$\N_\pm(\xi,\eta,w)$ are regular (holomorphic) on the cuts $L_+$
or $L_-$ respectively their suffices, and the algebraic
relations are satisfied on $L_\pm$:
\begin{equation}\label{LocalConstraints}
\begin{array}{lcllll}
L_+,:&&\k_+\cdot\N_+=0,&\M_+\cdot\l_+=0,&\bphi_+\cdot\M_+=0,&
\N_+\cdot\bpsi_+=0\\[2ex]
L_-\,:&&\k_-\cdot\N_-=0,&\M_-\cdot\l_-=0,&\bphi_-\cdot\M_-=0,&
\N_-\cdot\bpsi_-=0\end{array}
\end{equation}
The local structure (\ref{LocalStructure}) with the constraints
(\ref{LocalConstraints}) allow to clarify the monodromy
properties of $\bigpsi(\xi,\eta,w)$ near its branchpoints.
This structure is determined by the monodromy matrices
$\T_+$ or $\T_-$ characterizing the linear transformations
which  relate the matrix $\bigpsi(\xi,\eta,w)$ and its
analytical continuations along, say, the clockwise directed
paths $t_+$ and $t_-$ joining the left edge of the cut $L_+$ or
$L_-$ respectively with its right edge:
\begin{equation}\label{Tmatrices}
\bigpsi(\xi,\eta,w)\stackrel {t_\pm}
\longrightarrow\bigpsi(\xi,\eta,w)\cdot \T_\pm(w),\qquad
\T_\pm(w)={\bf I}-2 {\l_\pm(w)\otimes\k_\pm(w)\over
(\l_\pm(w)\cdot\k_\pm(w))}
\end{equation}
 For derivation of these
expressions, besides (\ref{LocalStructure}) and
(\ref{LocalConstraints}), we have used the properties
$\lambda_+\stackrel {t_+} \longrightarrow -\lambda_+$ and
$\lambda_-\stackrel {t_-} \longrightarrow -\lambda_-$. One can
observe easily that each of the monodromy matrices $\T_\pm(w)$
satisfies the identity  $\T_\pm^2(w)\equiv {\bf I}$.\footnote{
We stress the point, that these properties take place for
vacuum and electrovacuum cases only. These don't hold in the
presence of the Weyl spinor field, when the local structures
(\ref{LocalStructure}) become more complicate and the
branchpoints can be nonalgebraic
\cite{Alekseev:1985,Alekseev:2000a}.}

As it was shown in \cite{Alekseev:1985,Alekseev:1987}, the linear
relations between $\k^\dagger_\pm(w)$ and $\l_\pm(w)$  implied by
the conditions (\ref{W-equations}) allow to express the components of
the monodromy matrices (\ref{Tmatrices}) in terms of the
``projective'' vectors $\k_\pm(w)$. These relations and
the affine parametrization  for $\k_\pm(w)$ of the form
\begin{equation}\label{klExpressions}
\l_\pm(w)=4\epsilon_0(w-\xi_0) (w-\eta_0)\W_0^{-1}(w)\cdot
\k_\pm^\dagger(w),\quad \k_\pm(w)=\left\{1,\hskip1ex\u_\pm(w),
\hskip1ex\v_\pm(w)\right\}
\end{equation}
allow also to express all components of the monodromy matrices
$\T_\pm(w)$ in terms of four scalar functions $\u_\pm(w)$ and
$\v_\pm(w)$ which we call as the monodromy data. These data are
conserved, i.e. they are coordinate independent, and they can be
determined (at least in principle) for any local solution of the
reduced vacuum or electrovacuum field equations. For vacuum
$\v_\pm(w)\equiv 0$ and therefore, any vacuum solution is
characterized by two functions $\u_\pm(w)$ only.

\subsection{The inverse problem of the monodromy
transform}\label{InverseProblem}

Another important property of the nonevolving monodromy data is
that they characterize unambiguously every local solution. For a
given monodromy data the corresponding local solution can be
calculated in quadratures in terms of solution of the linear
singular integral equations which scalar kernels are constructed
using a given monodromy data. It is remarkable, that for any
choice of the monodromy data $\u_+(w)$, $\v_+(w)$ holomorphic
near the point $w=\xi_0$ and $\u_-(w)$, $\v_-(w)$ holomorphic
near the point $w=\eta_0$ the solution of these singular
integral equations always exists and it is unique. These
integral equations, solving the inverse problem of such
monodromy transform, are  equivalent to the
integrable reductions of Einstein equations.  In
\cite{Alekseev:2000a, Alekseev:2000b} these equations have been
presented in two alternative forms and together with their
equivalent regularizations.
Here we recall the singular forms of
these equations only:  \begin{equation}\label{SLINEs}
\displaystyle{1\over \pi i}\vpint\limits_L\,{ {\cal
K}(\tau,\zeta)\over
\zeta-\tau}\,{\bphi}(\xi,\eta,\zeta)\,d\,\zeta={\bf k}(\tau),\quad
\displaystyle{1\over \pi i}\vpint\limits_L\,{ \widetilde{\cal
K}(\zeta,\tau)\over
\zeta-\tau}\,{\bpsi}(\xi,\eta,\zeta)\,d\,\zeta={\bf
l}(\tau)
\end{equation}
with scalar kernels of the following
structures (the arguments $\xi$, $\eta$ are omitted):
\begin{equation}\label{Kernels}
{\cal K}(\tau,\zeta)=-[\lambda]_\zeta
(\k(\tau)\cdot\l(\zeta)),\qquad \widetilde{\cal
K}(\tau,\zeta)=-[\lambda^{-1}]_\zeta (\k(\zeta)\cdot\l(\tau))
\end{equation}
The coordinates $\xi$ and $\eta$ enter the integral equations
(\ref{SLINEs}) as parameters which determine the location of
the endpoints of the integration paths and as arguments of the
$\lambda$-multipliers in the kernels  (\ref{Kernels}).
The expressions $[\lambda]_\zeta$ in (\ref{Kernels}) mean  the
jump (i.e. a half of the difference between the values of the
function on the left and right edges of a cut) of the function
$\lambda$ at the point $\zeta\in L$.
It seems useful to recall here the way for derivation of the
basic equations (\ref{SLINEs}) which was suggested in
\cite{Alekseev:1985} and to explain with more details their
structure.

The described above general analytical properties of $\bigpsi$
and $\bigpsi^{-1}$ on the spectral plane permit to represent
these matrix functions as Cauchy integrals over the cut
$L=L_++L_-$ where the integrand densities are the
jumps of these matrix functions on $L$. These jumps are
represented by the first terms in the right hand sides of the
expressions (\ref{LocalStructure}), while the second terms there
represent the continuous parts of these integrals determined by
the principal values of the Cauchy integrals with the same
densities. The integral equations (\ref{SLINEs}) arise
immediately, if we use the integral representations mentioned
just above in the first two algebraic constraints in each line
of (\ref{LocalConstraints}).

To explain the structure of the integral equations (\ref{SLINEs}),
we recall that the integrals there are calculated over the cut
consisting of two disconnected parts:  $L=L_++L_-$, that each of
the Cauchy type integrals in (\ref{SLINEs}) splits into the sum
of two integrals over $L_+$ and $L_-$ respectively and only one
of these two integrals is a singular one. In the integrands
$[\lambda]_\zeta$ means the jump of the function
$\lambda_+(\xi,\eta,\zeta)$ if $\zeta=\zeta_+\in L_+$ or of the
function $\lambda_-(\xi,\eta,\zeta)$ if $\zeta=\zeta_-\in L_-$.
The unknown vector functions $\bphi(\xi,\eta,\tau)$,
$\bpsi(\xi,\eta,\tau)$ and the vector functions $\k(\tau)$,
$\l(\tau)$ in (\ref{SLINEs}) should get the suffix ``$+$'', if their
argument $\tau=\tau_+\in L_+$ and the suffix ``$-$'', if their
argument $\tau=\tau_-\in L_-$, and the corresponding suffixed
vector functions should be identified with the fragments of the
local structure of $\bigpsi$ defined in (\ref{LocalStructure}).
Because the parameter $\tau$ in the equations (\ref{SLINEs})
also run the entire cut $L$, i.e. it should take the values
$\tau=\tau_+\in L_+$ as well as $\tau=\tau_-\in L_-$, each of
the equations (\ref{SLINEs}) represents a coupled pair of vector
integral equations. More explicitly, the first of the equations
(\ref{SLINEs}) can be represented as a system $$
\begin{array}{l}\displaystyle{1\over \pi i}\vpint\limits_{L_+}
{{\cal K}(\tau_+,\zeta_+)\over
\zeta_+-\tau_+}\,{\bphi}_+(\zeta_+)\,d\,\zeta_+ +{1\over \pi
i}\int\limits_{L_-}{{\cal K}(\tau_+,\zeta_-)\over
\zeta_--\tau_+}\,{\bphi}_-(\zeta_-)\,d\,\zeta_-={\bf
k}_+(\tau_+)\\[1ex] \displaystyle{1\over \pi
i}\int\limits_{L_+}{{\cal K}(\tau_-,\zeta_+)\over
\zeta_+-\tau_-}\,{\bphi}_+(\zeta_+)\,d\,\zeta_+ +
\displaystyle{1\over \pi i}\vpint\limits_{L_-}{{\cal
K}(\tau_-,\zeta_-)\over
\zeta_--\tau_-}\,{\bphi}_-(\zeta_-)\,d\,\zeta_-={\bf
k}_-(\tau_-) \end{array}$$
with unknown vector functions $\bphi_+(\xi,\eta,\tau_+)$,
$\bphi_-(\xi,\eta,\tau_-)$ and the scalar kernels
$$\begin{array}{l} {\cal
K}(\tau_\pm,\zeta_+)=-[\lambda_+]_{\zeta_+}
(\k_\pm(\tau_\pm)\cdot \l_+(\zeta_+))\\[1ex] {\cal
K}(\tau_\pm,\zeta_-)=-[\lambda_-]_{\zeta_-}
(\k_\pm(\tau_\pm)\cdot \l_-(\zeta_-)).\end{array}$$
Similarly, the second equation in (\ref{SLINEs}) leads to an
equivalent system of integral equations for two unknown vector
functions $\bpsi_+(\xi,\eta,\tau_+)$,
$\bpsi_-(\xi,\eta,\tau_-)$:
$$\begin{array}{l}\displaystyle{1\over \pi i}\vpint\limits_{L_+}
{\widetilde{\cal K}(\tau_+,\zeta_+)\over
\zeta_+-\tau_+}\,{\bpsi}_+(\zeta_+)\,d\,\zeta_+ +{1\over \pi
i}\int\limits_{L_-}{\widetilde{\cal K}(\tau_+,\zeta_-)\over
\zeta_--\tau_+}\,{\bpsi}_-(\zeta_-)\,d\,\zeta_-=
\l_+(\tau_+)\\[1ex] \displaystyle{1\over \pi
i}\int\limits_{L_+}{\widetilde{\cal K}(\tau_-,\zeta_+)\over
\zeta_+-\tau_-}\,{\bpsi}_+(\zeta_+)\,d\,\zeta_+ +
\displaystyle{1\over \pi i}\vpint\limits_{L_-}{\widetilde{\cal
K}(\tau_-,\zeta_-)\over
\zeta_--\tau_-}\,{\bpsi}_-(\zeta_-)\,d\,\zeta_-= \l_-(\tau_-)
\end{array}$$
where the kernels $\widetilde{\cal K}_{\pm\pm}$ possess the
forms ``almost symmetric'' to that of ${\cal K}_{\pm\pm}$:
$$\begin{array}{l} \widetilde{\cal
K}(\tau_\pm,\zeta_+)=-[\lambda_+^{-1}]_{\zeta_+}
(\k_+(\zeta_+)\cdot \l_\pm(\tau_\pm))\\[1ex] \widetilde{\cal
K}(\tau_\pm,\zeta_-)=-[\lambda_-^{-1}]_{\zeta_-}
(\k_-(\zeta_-)\cdot \l_\pm(\tau_\pm)).\end{array}$$
Vector solutions $\bphi_+(\xi,\eta,\tau_+)$, $
\bphi_-(\xi,\eta,\tau_-)$ and $\bpsi_+(\xi,\eta,\tau_+)$,
$\bpsi_-(\xi,\eta,\tau_-)$ of these equations together with the
corresponding monodromy data vectors $\k_+(\tau_+)$,
$\k_-(\tau_-)$ and with expressions (\ref{klExpressions}) for
$\l_+(\tau_+)$, $ \l_-(\tau_-)$ determine the solution of our
spectral problem (\ref{UVW}) -- (\ref{Normalization})) by means
of the quadratures
$$\begin{array}{r} \bigpsi(\xi,\eta,w)={\bf
I} +\displaystyle{1\over \pi i}\int\limits_{L_+}\,{
[\lambda_+^{-1}]_{\zeta_+} \over \zeta_+-w}\,
\bpsi_+(\zeta_+)\otimes\k_+(\zeta_+)\,d\,\zeta_+\\[1ex]+
\displaystyle{1\over \pi i}\int\limits_{L_-}\,{
[\lambda_-^{-1}]_{\zeta_-} \over \zeta_--w}\,
\bpsi_-(\zeta_-)\otimes\k_-(\zeta_-)\,d\,\zeta_- \end{array}$$
$$\begin{array}{r} \bigpsi^{-1}(\xi,\eta,w)={\bf I}
+\displaystyle{1\over \pi i}\int\limits_{L_+}\,{
[\lambda_+]_{\zeta_+} \over \zeta_+-w}\,
\l_+(\zeta_+)\otimes\bphi_+(\zeta_+)\,d\,\zeta_+\\[1ex]+
\displaystyle{1\over \pi i}\int\limits_{L_-}\,{
[\lambda_-]_{\zeta_-} \over \zeta_--w}\,
\l_-(\zeta_-)\otimes\bphi_-(\zeta_-)\,d\,\zeta_- \end{array}$$

\subsection{Calculation of the field components and potentials}

All components of the solution can be expressed in terms of the
matrix $\R(\xi,\eta)$ determined by the
asymptotic expansions \cite{Alekseev:1987}
\begin{equation}\label{RDefinition}
\bigpsi=\mbox{\bf
I}+w^{-1}\R+O(w^{-2}), \quad \bigpsi^{-1}=\mbox{\bf
I}-w^{-1}\R+O(w^{-2})
\end{equation}
Hence, for this matrix we have the
 following alternative expressions:
$$\begin{array}{l}
\R=\displaystyle{1\over \pi i}\int\limits_{L_+}\,
[\lambda_+]_{\zeta_+}\,
\l_+(\zeta_+)\otimes\bphi_+(\zeta_+)\,d\,\zeta_++
\displaystyle{1\over \pi i}\int\limits_{L_-}\,
[\lambda_-]_{\zeta_-}\,
\l_-(\zeta_-)\otimes\bphi_-(\zeta_-)\,d\,\zeta_- \\[1ex]
=-\displaystyle{1\over \pi i}\int\limits_{L_+}
[\lambda_+^{-1}]_{\zeta_+}
\bpsi_+(\zeta_+)\otimes\k_+(\zeta_+)\,d\,\zeta_+-
\displaystyle{1\over \pi i}\int\limits_{L_-}
[\lambda_-^{-1}]_{\zeta_-}
\bpsi_-(\zeta_-)\otimes\k_-(\zeta_-)\,d\,\zeta_- \end{array}
$$
The matrices $\U$, $\V$, $\W$ and the Ernst
potentials then can be expressed as follows
\begin{equation}\label{SolutionA}
\begin{array}{lcccl} \U=2
i\partial_\xi\R,\quad \V=2 i\partial_\eta\R, &&&& {\cal
E}=\epsilon_0-2 i R_3{}^4\\[1ex] \W=\W_0(w)-4 i
(\bigomega\R+\R^\dagger\bigomega),&&&& {\Phi}=2 i R_3{}^5
\end{array}
\end{equation}
where the components $R_A{}^B$ of the $3\times
3$-matrix $\R$ are numbered by the index values
$A,B\ldots=3,4,5$.  The corresponding expressions for the metric
components $g_{ab}$ with $a,b,\ldots=3,4$ and nonzero
components of complex electromagnetic potential $\Phi_a$ are
\cite{Alekseev:1987}:
\begin{equation}\label{SolutionB}
\begin{array}{l}
g_{33}=\epsilon_0-i(R_3{}^4-\overline{R}_3{}^4)+
\Phi_3\overline{\Phi}_3,\\[1ex]
g_{34}=-i(\beta-\beta_0)+i(R_3{}^3+\overline{R}_4{}^4)+
\Phi_3\overline{\Phi}_4,\\[1ex]
g_{44}=\epsilon_0\epsilon\alpha_0^2+i(R_4{}^3-\overline{R}_4{}^3)+
\Phi_4\overline{\Phi}_4,
\end{array}\quad
\left(\begin{array}{l}
\Phi_3\\[1ex] \Phi_4\end{array}\right)=2 i \left(\begin{array}{l}
R_3{}^5\\[1ex] R_4{}^5\end{array}\right)
\end{equation}

\section{The ``integral evolution equations''}

In this section we present a new way for construction of integral
equations equivalent to the dynamical part of reduced Einstein
equations. As in the previous section we concentrate our
consideration on the vacuum Einstein equations and electrovacuum
Einstein - Maxwell equations, however this construction of the
integral equations can be realized for any of integrable reductions
of Einstein equations mentioned in the Introduction. The
structure of the derived here new equations called also as
``integral evolution equations'' differes essentially from the
structure of singular integral equations (\ref{SLINEs}) and
their simplest regularizations described in various forms in
\cite{Alekseev:2000a,Alekseev:2000b}.

\subsection{The ``in-states'' $\bigpsi_+(\xi,w)$ and
$\bigpsi_-(\eta,w)$}

As the first step we introduce two
particular values $\bigpsi_+(\xi,w)$ and $\bigpsi_-(\eta,w)$ of
the matrix function $\bigpsi(\xi,\eta,w)$. In view of some
analogy  with the scattering problem we call them as
``in-states''. These ``in-states'' are defined as
\begin{equation}\label{InStates}
\bigpsi_+(\xi,w)=\bigpsi(\xi,\eta_0,w),\qquad
\bigpsi_-(\eta,w)=\bigpsi(\xi_0,\eta,w).
\end{equation}
For the hyperbolic case these are the boundary values of
$\bigpsi(\xi,\eta,w)$ on the characteristics which pass through the
reference point $P_0(\xi_0,\eta_0)$ in the orbit space, while
for the elliptic case $\bigpsi_+$ and $\bigpsi_-$ are the
limit values, when $\eta\to\eta_0$ or $\xi\to\xi_0$
respectively, of such analytical extension of the matrix
function $\bigpsi(\xi,\eta,w)$ which arguments $\xi$ and $\eta$
are independent complex variables instead of being complex
conjugated to each other. The matrix functions (\ref{InStates})
can be defined also as the normalized fundamental solutions of
the linear ordinary systems which are the restrictions of the
system (\ref{UV-equations}) to the corresponding characteristics
$\eta=\eta_0$ and $\xi=\xi_0$ in the orbit space (for the hyperbolic
case) or the restrictions of the analytically extended system
(\ref{UV-equations}) to the complex surfaces $\eta=\eta_0$ and
$\xi=\xi_0$ in the complexified orbit space (for the elliptic case):
$$\left\{\begin{array}{l}
2i(w-\xi)\partial_\xi\bigpsi_+=\U(\xi,\eta_0)\cdot
\bigpsi_+\\[1ex]
\bigpsi_+(\xi_0,w)={\bf I}\end{array}\right.\qquad
\left\{\begin{array}{l}
2i(w-\eta)\partial_\eta\bigpsi_-=\V(\xi_0,\eta)\cdot
\bigpsi_-\\[1ex]
\bigpsi_-(\eta_0,w)={\bf I}\end{array}\right.$$
Therefore, the function $\bigpsi_+(\xi,w)$ and its inverse are
holomorphic outside $L_+$, while $\bigpsi_-(\eta,w)$ and its
inverse are holomorphic outside $L_-$ and each of these matrix
functions  posses only two brachpoints on the spectral plane,
and these branchpoints coincide with the endpoints of
the cuts $L_+$ and $L_-$ respectively. On these
cuts $\bigpsi_\pm$ possess the structures similar to
(\ref{LocalStructure}):
\begin{equation}\label{InitialStructure}
\begin{array}{ll}
L_+:&\bigpsi_+(\xi,w)=
\lambda_+^{-1}\bpsi_{0+}(\xi,w)\otimes\k_+(w)+
\M_{0+}(\xi,w)\\[2ex]
&\bigpsi_+^{-1}(\xi,w)=
\lambda_+\l_+(w)\otimes\bphi_{0+}(\xi,w)+\N_{0+}(\xi,w)\\[2ex]
L_-:&\bigpsi_-(\eta,w)=
\lambda_-^{-1}\bpsi_{0-}(\eta,w)\otimes\k_-(w)+
\M_{0-}(\eta,w)\\[2ex]
&\bigpsi_-^{-1}(\eta,w)=
\lambda_-\l_-(w)\otimes\bphi_{0-}(\eta,w)+\N_{0-}(\eta,w)
\end{array}
\end{equation}
where $\lambda_+$ and $\lambda_-$, $\k_\pm(w)$ and $\l_\pm(w)$ are
the same as in (\ref{LocalStructure}); each of the row and column
vectors $\bpsi_{0\pm}$, $\bphi_{0\pm}$ and
the matrices $\M_{0\pm}$, $\N_{0\pm})$
are regular (holomorphic) on the cuts $L_+$ or $L_-$
respectively their suffices, and the following algebraic
constraints on the fragments of these local structures similar
to (\ref{LocalConstraints}) are satisfied on the cuts $L_\pm$:
$$\begin{array}{lcllll}
L_+\,:&&\k_+\cdot\N_{0+}=0,&\M_{0+}\cdot\l_+=0,&\bphi_{0+}
\cdot\M_{0+}=0,&
\N_{0+}\cdot\bpsi_{0+}=0\\[2ex]
L_-\,:&&\k_-\cdot\N_{0-}=0,&\M_{0-}\cdot\l_-=0,&\bphi_{0-} \cdot
\M_{0-}=0,&\N_{0-}\cdot\bpsi_{0-}=0\end{array}
$$

\subsection{``Scattering'' matrices $\bchi_\pm(\xi,\eta,w)$ and
``dynamical'' monodromy data $\m_+(\eta,w)$ and $\m_-(\xi,w)$}
For the next step of our construction we introduce the
``dressing'' or ``scattering'' matrices $\bchi_\pm(\xi,\eta,w)$
presenting $\bigpsi(\xi,\eta,w)$ in two alternative forms
\begin{equation}\label{Dressing}
\begin{array}{l}
\bigpsi(\xi,\eta,w)=\bchi_+(\xi,\eta,w)\cdot \bigpsi_+(\xi,w)\\[1ex]
\bigpsi(\xi,\eta,w)=\bchi_-(\xi,\eta,w)\cdot
\bigpsi_-(\eta,w)\end{array}
\end{equation}
The basic idea of such alternative
representation is based on the fact that the monodromy properties of
$\bigpsi(\xi,\eta,w)$ on the spectral plane $w$ are conserved
during the evolution of the fields prescribed by the reduced
Einstein equations. Therefore, the monodromy properties of
$\bigpsi(\xi,\eta,w)$  should be the same as for $\bigpsi_+(\xi,w)$
near the cut $L_+$ and the same as for $\bigpsi_-(\eta,w)$ near the
cut $L_-$. This allows to conjecture that the matrix function
$\bchi_+(\xi,\eta,w)$ should be regular on the cut $L_+$ and
$\bchi_-(\xi,\eta,w)$ should be regular on the cut $L_-$. And, we
have all what is necessary to check this conjecture. Namely, we know
the analytical properties of $\bigpsi$ and $\bigpsi_\pm$ on the
spectral plane and therefore, the structures of
$\bchi_\pm(\xi,\eta,w)$ on the spectral plane $w$ can be described
in details. In particular, it is easy to see that each of the
matrices $\bchi_\pm(\xi,\eta,w)$ is holomorphic function
everywhere outside the cut $L=L_++L_-$. Using the expressions
$$\bchi_+(\xi,\eta,w)\equiv\bigpsi(\xi,\eta,w)\cdot
\bigpsi_+^{-1}(\xi,\eta_0,w),\quad
\bchi_-(\xi,\eta,w)\equiv\bigpsi(\xi,\eta,w)\cdot
\bigpsi_-^{-1}(\xi_0,\eta,w)$$
we get the local structures of
$\bchi_\pm$ and $\bchi_\pm^{-1}$ on $L_+$ in the forms
$$\begin{array}{ll} \bchi_+=\left(\k_+(w)\cdot\l_+(w)\right)
\bpsi_+(\xi,\eta,w)\otimes\bphi_{0+}(\xi,w)+
\M_+(\xi,\eta,w)\cdot\N_{0+}(\xi,w)\\[1ex]
\bchi_+^{-1}= \left(\k_+(w)\cdot\l_+(w)\right)
\bpsi_{+0}(\xi,w)\otimes\bphi_+(\xi,\eta,w)+
\M_{0+}(\xi,w)\cdot\N_+(\xi,\eta,w)\\[1ex]
\bchi_-=\left(\lambda_+^{-1}\bpsi_+(\xi,\eta,w)\otimes\k_+(w)+
\M_+(\xi,\eta,w)\right)\cdot\bigpsi_-^{-1}(\eta,w)\\[1ex]
\bchi_-^{-1}= \bigpsi_-(\eta,w)\cdot
\left(\lambda_+\l_+(w)\otimes\bphi_+(\xi,\eta,w)+\N_+(\xi,\eta,w)
\right)\end{array}$$
and on $L_-$ in the similar forms:
$$\begin{array}{l}
\bchi_+=\left(\lambda_-^{-1}\bpsi_-(\xi,\eta,w)\otimes\k_-(w)+
\M_-(\xi,\eta,w)
\right)\cdot\bigpsi_+^{-1}(\xi,w)\\[1ex]
\bchi_+^{-1}=\bigpsi_+(\xi,w)\cdot
\left(\lambda_-\l_-(w)\otimes\bphi_-(\xi,\eta,w)+\N_-(\xi,\eta,w)
\right)\\[1ex]
\bchi_-=\left(\k_-(w)\cdot\l_-(w)\right)
\bpsi_-(\xi,\eta,w)\otimes\bphi_{0-}(\eta,w)+
\M_-(\xi,\eta,w)\cdot\N_{0-}(\eta,w)\\[1ex]
\bchi_-^{-1}= \left(\k_-(w)\cdot\l_-(w)\right)
\bpsi_{0-}(\eta,w)\otimes\bphi_-(\xi,\eta,w)+
\M_{0-}(\eta,w)\cdot\N_-(\xi,\eta,w)\end{array}$$
As it was
expected, the functions $\bchi_+$, $\bchi_+^{-1}$ and $\bchi_-$,
$\bchi_-^{-1}$ turn out to be regular on $L_+$ and $L_-$
respectively. These expressions show also that $\bchi_+$ and its
inverse possess the jumps on $L_-$ and $\bchi_-$ and its inverse
possess the jumps on $L_+$. These jumps are degenerate matrices
which rank is equal to $1$ and which therefore, can be represented
as the products of column and row - vectors
\begin{equation}\label{chiJumps}
\begin{array}{l}
[\bchi_+]_{L_-}=[\lambda_-^{-1}]\bpsi_-(\xi,\eta,\tau_-)\otimes
\m_-(\xi,\tau_-)\\[1ex]
[\bchi_-]_{L_+}= [\lambda_+^{-1}]\bpsi_+(\xi,\eta,\tau_+)\otimes
\m_+(\eta,\tau_+)\\[2ex]
[\bchi_+^{-1}]_{L_-}=[\lambda_-] \p_-(\xi,\tau_-)\otimes
\bphi_-(\xi,\eta,\tau_-)\\[1ex]
[\bchi_-^{-1}]_{L_+}=[\lambda_+]\p_+(\eta,\tau_+)\otimes
\bphi_+(\xi,\eta,\tau_+)\end{array}
\end{equation}
where $\tau_+\in L_+$ and $\tau_-\in L_-$. Here it is important
that the vector-functions $\m_-(u,w)$, $\p_-(u,w)$ and
$\m_+(\eta,w)$, $\p_+(\eta,w)$ are dependent (besides the
spectral parameter $w=\tau_+$ on $L_+$ or $w=\tau_-$ on $L_-$)
upon only one of two coordinates $\xi$ and $\eta$ respectively.
These vector functions are determined by the expressions:
\begin{equation}\label{mpvectors}
\begin{array}{lcl}
\m_-(\xi,w)=\k_-(w)\cdot\bigpsi_+^{-1}(\xi,w) &&
\p_-(\xi,w)=\bigpsi_+(\xi,w)\cdot\l_-(w)\\[1ex]
\m_+(\eta,w)=\k_+(w)\cdot\bigpsi_-^{-1}(\eta,w)&&
\p_+(\eta,w)=\bigpsi_-(\eta,w)\cdot\l_+(w)
\end{array}
\end{equation}
The components of these vector functions are determined completely by
the ``in-states'' $\bigpsi_+(\xi,w)$ and $\bigpsi_-(\eta,w)$. It can
be recalled here that the components of the conserved monodromy data
vectors $\k_\pm(w)$ and $\l_\pm(w)$ in turn are determined by
$\bigpsi_\pm$ as certain fragments of the local structures
(\ref{InitialStructure}) of these matrix functions. It is useful
to note here also that the vector functions (\ref{mpvectors})
can be interpreted as the monodromy data for the
scattering matrices $\bchi_\pm$ on the spectral plane. To show
this we consider the monodromy matrices $\widetilde{\T}_\pm$
which relate, similarly to (\ref{Tmatrices}), the values of
$\bchi_\pm$ with their analytical continuations along the paths
$t_\pm$ surrounding the branchpoints $w=\xi$ and $w=\eta$ in the
clockwise directions:
$$\bchi_-(\xi,\eta,w)\stackrel {t_+}
\longrightarrow\bchi_-(\xi,\eta,w)\cdot
\widetilde{\T}_+(\eta,w),\quad \bchi_+(\xi,\eta,w)\stackrel {t_-}
\longrightarrow\bchi_+(\xi,\eta,w)\cdot \widetilde{\T}_-(\xi,w)$$
The described earlier local structures of $\bchi_\pm$ on the cuts
$L_\pm$ permit to derive the following expressions for these new
monodromy matrices
$$\begin{array}{l}\widetilde{\T}_+(\eta,w)=\bigpsi_-(\eta,w)
\cdot\T_+(w)\cdot\bigpsi_-^{-1}(\eta,w)= {\bf I}-2
\displaystyle{\p_+(\eta,w)\otimes\m_+(\eta,w)\over
(\p_+(\eta,w)\cdot\m_+(\eta,w))},\\[3ex]
\widetilde{\T}_-(\xi,w)=\bigpsi_+(\xi,w)\cdot\T_-(w)\cdot
\bigpsi_+^{-1}(\xi,w)= {\bf I}-2
\displaystyle{\p_-(\xi,w)\otimes\m_-(\xi,w)\over
(\p_-(\xi,w)\cdot\m_-(\xi,w))},\end{array} $$
where we have used
again the properties of the functions $\lambda_\pm$, that
$\lambda_+\stackrel {t_+} \longrightarrow -\lambda_+$ and
$\lambda_-\stackrel {t_-} \longrightarrow -\lambda_-$. Unlike
(\ref{Tmatrices}) the matrices $\widetilde{\T}_\pm$ possess some
evolution (i.e. a coordinate dependence), however this
evolution do not violate their property to satisfy identically
the relations $\widetilde{\T}_\pm^2={\bf I}$, which
characterizes the conservation of the simple algebraic character
of the branchpoints of $\bchi_\pm$.  (In the presence of a Weyl
spinor field the character of the branching and therefore, the
expressions for the monodromy matrices $\widetilde{\T}_\pm$
change and the identities mentioned above do not take place.
Thus, we see that the ``projective'' vectors $\m_+(\eta,w)$,
$\m_-(\xi,w)$ and $\p_+(\eta,w)$, $\p_-(\xi,w)$ play a role of
some evolving analog of the conserved monodromy data (or the
scattering data) $\k_\pm(w)$, $\l_\pm(w)$ for the solutions of
the system (1), and hence, we call them as ``dynamical monodromy
data''.

\subsection{Consistency of representations (\ref{Dressing}) for
$\bigpsi(\xi,\eta,w)$}

Earlier it was shown, that the matrix
function $\bigpsi(\xi,\eta,w)$ is holomorphic on the $w$-plane
outside the cut $L_++ L_-$.  Similarly, $\bigpsi_+$ is
holomorphic on the entire $w$-plane outside the cut $L_+$ and
$\bigpsi_-$ is holomorphic on the entire $w$-plane outside the
cut $L_-$, and the same is true also for the inverse matrices
respectively.  These properties and the local structures
(\ref{LocalStructure}) and (\ref{InitialStructure}) allowed us
to conclude in the previous subsection that the function
$\bchi_+(\xi,\eta,w)$ is regular on $L_+$ and possesses a jump
on $L_-$, while the function $\bchi_-(\xi,\eta,w)$ is regular on
$L_-$ and possesses a jump on $L_+$. It is easy to see also that
these functions should satisfy the conditions
$$\bchi_+(\xi,\eta,w=\infty)=\hbox{\bf
I},\qquad\bchi_-(\xi,\eta,w=\infty)=\hbox{\bf I}$$
This means, that we can represent $\bchi_+$ and $\bchi_-$ as the
integrals over $L_-$ and over $L_+$ respectively:
$$\bchi_+(\xi,\eta,w)=\hbox{\bf I}+\displaystyle{1\over i\pi}
\int\limits_{L_-}\displaystyle{[\bchi_+]_{\zeta_-}\over\zeta_--w}
\,d\zeta_-\qquad \bchi_-(\xi,\eta,w)=\hbox{\bf
I}+\displaystyle{1\over i\pi}
\int\limits_{L_+}\displaystyle{[\bchi_-]_{\zeta_+}\over\zeta_+-w}
\,d\zeta_+$$
Substituting here the expressions
(\ref{chiJumps}) for the jumps $[\bchi_\pm]_{L_\mp}$ we get the
following representations of these matrices
\begin{equation}\label{chiIntegrals}
\begin{array}{l} \bchi_+(\xi,\eta,w)=\hbox{\bf
I}+\displaystyle{1\over i\pi} \int\limits_{L_-}\displaystyle{
[\lambda_-^{-1}]_{\zeta_-}\over\zeta_--w}
\bpsi_-(\xi,\eta,\zeta_-)\otimes \m_-(\xi,\zeta_-) \,d\zeta_-\\
\bchi_-(\xi,\eta,w)=\hbox{\bf I}+\displaystyle{1\over i\pi}
\int\limits_{L_+}\displaystyle{
[\lambda_+^{-1}]_{\zeta_+}\over\zeta_+-w}
\bpsi_+(\xi,\eta,\zeta_+)\otimes \m_+(\eta,\zeta_+)
\,d\zeta_+\end{array}
\end{equation}
Similar integral representations can be derived for the inverse
matrices:
\begin{equation}\label{InversechiIntegrals}
\begin{array}{l}
\bchi_+^{-1}(\xi,\eta,w)=\mbox{\bf I}+\displaystyle{1\over i\pi}
\int\limits_{L_-}\displaystyle{ [\lambda_-]_{\zeta_-}\over\zeta_--w}
\p_-(\xi,\zeta_-)\otimes\bphi_-(\xi,\eta,\zeta_-) \,d\zeta_-\\
\bchi_-^{-1}(\xi,\eta,w)=\mbox{\bf I}+\displaystyle{1\over i\pi}
\int\limits_{L_+}\displaystyle{ [\lambda_+]_{\zeta_+}\over\zeta_+-w}
\p_+(\eta,\zeta_+)\otimes\bphi_+(\xi,\eta,\zeta_+)
\,d\zeta_+\end{array}
\end{equation}
From the local representations
(\ref{InitialStructure}) for $\tau_+\in L_+$ and
$\tau_-\in L_-$ we get:
\begin{equation}\label{InitialJumps}
\begin{array}{lcl}
L_+:&&[\bigpsi_+]_{\tau_+}=[\lambda_+^{-1}]_{\tau_+}\bpsi_{0+}
(\xi,\tau_+)
\otimes\k_+(\tau_+),\\[1ex]
&&[\bigpsi_+^{-1}]_{\tau_+}=[\lambda_+]_{\tau_+}\l_+(\tau_+)\otimes
\bphi_{0+} (\xi,\tau_+),\\[2ex]
L_-:&&
[\bigpsi_-]_{\tau_-}=[\lambda_-^{-1}]_{\tau_-}\bpsi_{0-}(\eta,\tau_-)
\otimes\k_-(\tau_-),\\[1ex]&&
[\bigpsi_-^{-1}]_{\tau_-}=[\lambda_-]_{\tau_-}\l_-(\tau_-)\otimes
\bphi_{0-}(\eta,\tau_-).\end{array}
\end{equation}

Now we return to (\ref{Dressing}) and consider a
condition of consistency of these alternative expressions for
$\bigpsi$. This means that we have to satisfy the condition
which we present in two equivalent forms
\begin{equation}\label{Consistency}
\begin{array}{l}\bchi_+(\xi,\eta,w)\cdot \bigpsi_+(\xi,w)=
\bchi_-(\xi,\eta,w)\cdot \bigpsi_-(\eta,w)\\[2ex]
\bigpsi_+^{-1}(\xi,w)\cdot\bchi_+^{-1}(\xi,\eta,w) =
\bigpsi_-^{-1}(\eta,w)\cdot\bchi_-^{-1}(\xi,\eta,w)
\end{array}
\end{equation}
These conditions can be reduced considerably,
because both sides of each of them are analytical (holomorphic)
functions of the spectral parameter $w$ everywhere on the spectral
plane outside the composed cut $L=L_++ L_-$ and both sides
take the same value at $w=\infty$ which is a unit matrix.
Therefore, the conditions (\ref{Consistency}) are equivalent to
the condition that the jumps of the left and right hand sides of
each of the equations (\ref{Consistency}) on $L_+$ as well as on
$L_-$ should coincide:  \begin{equation}\label{Jumps}
\begin{array}{lcl}
L_+:&& \bchi_+(\tau_+)\cdot
[\bigpsi_+]_{\tau_+}= [\bchi_-]_{\tau_+}\cdot\bigpsi_-(\tau_+)\\[1ex]
&&[\bigpsi_+^{-1}]_{\tau_+}\cdot\bchi_+^{-1}(\tau_+)=
\bigpsi_-^{-1}(\tau_+)\cdot
[\bchi_-^{-1}]_{\tau_+}\\[2ex]
L_-:&&\bchi_-(\tau_-)\cdot [\bigpsi_-]_{\tau_-}=
[\bchi_+]_{\tau_-}\cdot\bigpsi_+(\tau_-)\\[1ex]
&&[\bigpsi_-^{-1}]_{\tau_-}\cdot\bchi_-^{-1}(\tau_-)=
\bigpsi_+^{-1}(\tau_-)\cdot[\bchi_+^{-1}]_{\tau_-}
\end{array}
\end{equation}

\subsection{Coupled systems of ``integral
evolution equations''}
Using  of the expressions
(\ref{chiIntegrals})-- (\ref{InitialJumps}) in (\ref{Jumps})
leads to the following coupled pairs of the linear integral
equations for different fragments of the local structures
(\ref{LocalStructure}). One of them is a system for the vector
functions $\bpsi_\pm(\xi,\eta,w)$
\begin{equation}\label{psiEquations}
\left\{\begin{array}{l} \bpsi_+(\xi,\eta,\tau_+)-
\displaystyle{\int\limits_{L_-}} {\cal S}_+(\xi,\eta,\tau_+,\zeta_-)
\bpsi_-(\xi,\eta,\zeta_-) \,d\zeta_-=
\bpsi_{0+}(\xi,\tau_+)\\[2ex]
\bpsi_-(\xi,\eta,\tau_-)- \displaystyle{\int\limits_{L_+}} {\cal
S}_-(\xi,\eta,\tau_-,\zeta_+) \bpsi_+(\xi,\eta,\zeta_+) \,d\zeta_+=
\bpsi_{0-}(\eta,\tau_-)\end{array}\right.
\end{equation}
where $\tau_+,\zeta_+\in L_+$ and $\tau_-,\zeta_-\in L_-$, and
the scalar kernels ${\cal S}_+$ and ${\cal S}_-$ are:
$$\begin{array}{l}
{\cal S}_+(\xi,\eta,\tau_+,\zeta_-)=\displaystyle{1\over i\pi}
\displaystyle{[\lambda_-^{-1}]_{\zeta_-}
\over\zeta_--\tau_+}\left(\m_-(\xi,\zeta_-)\cdot
\bpsi_{0+}(\xi,\tau_+)\right)\\
{\cal S}_-(\xi,\eta,\tau_-,\zeta_+)=\displaystyle{1\over i\pi}
\displaystyle{[\lambda_+^{-1}]_{\zeta_+}
\over\zeta_+-\tau_-}\left(\m_+(\eta,\zeta_+)\cdot
\bpsi_{0-}(\eta,\tau_-)\right)\end{array}$$
Another (alternative) system of equations which possess the
vector functions $\bphi_\pm(\xi,\eta,w)$ as unknown variables
takes the form \begin{equation}\label{phiEquations}
\left\{\begin{array}{l} \bphi_+(\xi,\eta,\tau_+)-
\displaystyle{\int\limits_{L_-}}
\widetilde{S}_+(\xi,\eta,\tau_+,\zeta_-) \bphi_-(\xi,\eta,\zeta_-)
\,d\zeta_-=
\bphi_{0+}(\xi,\tau_+)\\[2ex]
\bphi_-(\xi,\eta,\tau_-)- \displaystyle{\int\limits_{L_+}}
\widetilde{S}_-(\xi,\eta,\tau_-,\zeta_+) \bphi_+(\xi,\eta,\zeta_+)
\,d\zeta_+= \bphi_{0-}(\eta,\tau_-)\end{array}\right.
\end{equation}
where the scalar kernels $\widetilde{S}_+$ and $\widetilde{S}_-$ are:
$$\begin{array}{l}
\widetilde{S}_+(\xi,\eta,\tau_+,\zeta_-)=\displaystyle{1\over i\pi}
\displaystyle{[\lambda_-]_{\zeta_-}
\over\zeta_--\tau_+}\left(\bphi_{0+}(\xi,\tau_+)\cdot\p_-(\xi,\zeta_-)
\right)\\
\widetilde{S}_-(\xi,\eta,\tau_-,\zeta_+)=\displaystyle{1\over i\pi}
\displaystyle{[\lambda_+]_{\zeta_+}
\over\zeta_+-\tau_-}\left(\bphi_{0-}(\eta,\tau_-)\cdot\p_+(\eta,\zeta_+)
\right)\end{array}$$

The structures of the derived equations (\ref{psiEquations}) or
(\ref{phiEquations}) may seem to be simple enough, but
another form of these equations can be found useful as well.

\subsection{Decoupled ``integral evolution equations''}

A substitution of $\bpsi_-$ from the second of the
equations (\ref{psiEquations}) into the first and substitution
of $\bpsi_+$ from the first of the equations (\ref{psiEquations})
into the second leads to a pair of decoupled equations
\begin{equation}\label{psiEqs}
\begin{array}{l}\bpsi_+(\tau_+)-\displaystyle{\int
\limits_{L_+}}{\cal F}_+(\tau_+,\zeta_+)\bpsi_+(\zeta_+)
\,d\zeta_+=\f_+(\tau_+)\\[2ex]
\bpsi_-(\tau_-)-\displaystyle{\int \limits_{L_-}}{\cal
F}_-(\tau_-,\zeta_-)\bpsi_-(\zeta_-)\,d\zeta_-=\f_-(\tau_-)
\end{array}
\end{equation}
where the dependence of the kernels, the right hand
sides and the unknown functions on $\xi$ and $\eta$ was omitted for
brevity. The kernels and right hand sides of (\ref{psiEqs})
possess more complicate than in  (\ref{psiEquations}) and
(\ref{phiEquations}) structures:
$$\begin{array}{l}
{\cal F}_+(\tau_+,\zeta_+)=\displaystyle{\int\limits_{L_-}} {\cal
S}_+(\tau_+,\chi_-){\cal S}_-(\chi_-,\zeta_+)\, d\chi_-,\\[2ex]
{\cal F}_-(\tau_-,\zeta_-)=\displaystyle{\int\limits_{L_+}} {\cal
S}_-(\tau_-,\chi_+){\cal S}_+(\chi_+,\zeta_-)\, d\chi_+,\\[2ex]
\f_+(\tau_+)=\bpsi_{0+}(\tau_+)+\displaystyle{\int\limits_{L_-}}
{\cal S}_+(\tau_+,\chi_-)\bpsi_{0-}(\chi_-)\,d\chi_-\\[2ex]
\f_-(\tau_-)=\bpsi_{0-}(\tau_-)+\displaystyle{\int\limits_{L_+}}
{\cal S}_-(\tau_-,\chi_+)\bpsi_{0+}(\chi_+)\,d\chi_+
\end{array}$$
Similarly, we arrive at decoupled
equations for the vector functions $\bphi_\pm$:
\begin{equation}\label{phiEqs}
\begin{array}{l}\bphi_+(\tau_+)-\displaystyle{\int
\limits_{L_+}}{\cal G}_+(\tau_+,\zeta_+)\bphi_+(\zeta_+)
\,d\zeta_+={\g}_+(\tau_+)\\[2ex]
\bphi_-(\tau_-)-\displaystyle{\int \limits_{L_-}}{\cal
G}_-(\tau_-,\zeta_-)\bphi_-(\zeta_-)\,d\zeta_-={\g}_-(\tau_-)
\end{array}
\end{equation}
which kernels and right hand sides possess the expressions
$$\begin{array}{l}
{\cal G}_+(\tau_+,\zeta_+)=\displaystyle{\int\limits_{L_-}}
\widetilde{\cal S}_+(\tau_+,\chi_-)\widetilde{\cal
S}_-(\chi_-,\zeta_+)\, d\chi_-,\\[2ex] {\cal
G}_-(\tau_-,\zeta_-)=\displaystyle{\int\limits_{L_+}}
\widetilde{\cal S}_-(\tau_-,\chi_+)\widetilde{\cal
S}_+(\chi_+,\zeta_-)\, d\chi_+,\\[2ex]
\g_+(\tau_+)=\bphi_{0+}(\tau_+)+\displaystyle{\int\limits_{L_-}}
\widetilde{\cal
S}_+(\tau_+,\chi_-)\bphi_{0-}(\chi_-)\,d\chi_-\\[2ex]
\g_-(\tau_-)=\bphi_{0-}(\tau_-)+\displaystyle{\int\limits_{L_+}}
\widetilde{\cal S}_-(\tau_-,\chi_+)\bphi_{0+}(\chi_+)\,d\chi_+
\end{array}$$

For construction of any solution of reduced Einstein equations
it is sufficient to solve only one of the systems of coupled
vector integral equations (\ref{psiEquations}) or
(\ref{phiEquations}) or their decoupled vector forms
(\ref{psiEqs}) or (\ref{phiEqs}). Besides that, all of the
equations derived above, being vector equations, have scalar
kernels and hence, they decouple also into separate equations
for each of the vector component.  In terms of their solutions
all components of the solution of Einstein equations can be
determined in quadratures.

\subsection{Calculation of solution components}

In section \ref{InverseProblem} we recall the expressions
for the metric components $g_{ab}$
($a,b,\ldots=3,4$), the nonzero components of a complex
electromagnetic potential $\Phi_a$ as well as the matrices $\U$,
$\V$, $\W$ and the Ernst potentials for any solution of reduced
Einstein or Einstein - Maxwell equations in terms of the
components $R_A{}^B$ ($A,B,\ldots$) of the matrix $\R$ defined
by the asymptotic expansion (\ref{RDefinition}) of $\bigpsi$ for
$w\to\infty$ (see (\ref{SolutionA}), (\ref{SolutionB}) and
\cite{Alekseev:1987}). Here we present the expressions for these
components of solution in terms of the matrices
$\R_\pm(\xi,\eta)$ defined by the asymptotic expansions of the
scattering matrices
$$\bchi_\pm=\mbox{\bf
I}+w^{-1}\R_\pm+O(w^{-2}),\qquad \bchi_\pm^{-1}=\mbox{\bf
I}-w^{-1}\R_\pm+O(w^{-2})$$
In particular, for the matrices $\U$, $\V$ and $\W$ the
following expressions can be derived easily from asymptotic
considerations:
$$\begin{array}{ll}
\U(\xi,\eta)=\U(\xi,\eta_0)+2 i\partial_\xi\R_+,
& \W(\xi,\eta,w)=\W(\xi,\eta_0,w)-4 i
(\bigomega\R_++\R_+^\dagger\bigomega)\\[1ex]
\V(\xi,\eta)=\V(\xi_0,\eta)+2 i\partial_\eta\R_-,
& \phantom{\W(\xi,\eta,w)}=\W(\xi_0,\eta,w)-4 i
(\bigomega\R_-+\R_-^\dagger\bigomega)
\end{array}$$
For calculation of other components of the solution one can use
the expressions (\ref{SolutionB}) where the matrix $\R$ defined
in (\ref{RDefinition}) is expressed in terms of the matrices
$\R_\pm$. This last expression can be presented in two
alternative forms:
$$\R(\xi,\eta)=\R_+(\xi,\eta)+\R_-(\xi,\eta_0)=
\R_-(\xi,\eta)+\R_+(\xi_0,\eta). $$
To complete the present construction
we present the alternative expressions for $\R_\pm$
which follow from the asymptotic expansions of the integral
representations (\ref{chiIntegrals}) and
(\ref{InversechiIntegrals}):  $$\begin{array}{l} \R_+=-\displaystyle{1\over \pi
i}\int\limits_{L_-}\, [\lambda_-^{-1}]_{\zeta_-}\,
\bpsi_-(\xi,\eta,\zeta_-)\otimes\m_-(\xi,\zeta_-)\,d\,\zeta_-\\[1ex]
\phantom{ \R_+}=\displaystyle{1\over \pi i}\int\limits_{L_-}\,
[\lambda_-]_{\zeta_-}\,
\p_-(\xi,\zeta_-)\otimes\bphi_-(\xi,\eta,\zeta_-)\,d\,\zeta_-\\[2ex]
\R_-=-\displaystyle{1\over \pi i}\int\limits_{L_+}\,
[\lambda_+^{-1}]_{\zeta_+}\,
\bpsi_+(\xi,\eta,\zeta_+)\otimes\m_+(\eta,\zeta_+)\,d\,\zeta_+\\[1ex]
\phantom{ \R_-}=\displaystyle{1\over \pi i}\int\limits_{L_+}\,
[\lambda_+]_{\zeta_+}\,
\p_+(\eta,\zeta_+)\otimes\bphi_+(\xi,\eta,\zeta_+)\,d\,\zeta_+
\end{array}$$
These expressions allow to calculate all components of solutions
also in the alternative forms. For example, for the Ernst
potentials we have $$\begin{array}{l} {\cal E}(\xi,\eta)={\cal
E}(\xi,\eta_0)+\displaystyle{2\over \pi}
\displaystyle{\int\limits_{L_-}}[\lambda_-^{-1}]_{\zeta_-}
(\e_1\cdot\bpsi_-(\zeta_-))\,\,(\m_-(\zeta_-)\cdot\e_2)
\,d\zeta_-=\\[1ex]
\phantom{{\cal E}(\xi,\eta_0)}={\cal
E}(\xi_0,\eta)+\displaystyle{2\over \pi}
\int\limits_{L_+}[\lambda_+^{-1}]_{\zeta_+}
(\e_1\cdot\bpsi_+(\zeta_+))\,\,(\m_+(\zeta_+)\cdot\e_2)
\,d\zeta_-\\[2ex]
\Phi(\xi,\eta)=\Phi(\xi,\eta_0)-\displaystyle{2\over \pi}
\int\limits_{L_-}[\lambda_-^{-1}]_{\zeta_-}
(\e_1\cdot\bpsi_-(\zeta_-))\,\,(\m_-(\zeta_-)\cdot\e_3)
\,d\zeta_-=\\[1ex]
\phantom{\Phi(\xi,\eta)}=\Phi(\xi_0,\eta)-\displaystyle{2\over \pi}
\int\limits_{L_+}[\lambda_+^{-1}]_{\zeta_+}
(\e_1\cdot\bpsi_+(\zeta_+))\,\,(\m_+(\zeta_+)\cdot\e_3) \,d\zeta_-
\end{array}$$
where $\e_1=\{1,0,0\}$, $\e_2=\{0,1,0\}$ and $\e_3=\{0,0,1\}$.

\section{Concluding remarks}
In this paper the two-dimensional space-time symmetry
reductions of vacuum Einstein equations and electrovacuum
Einstein - Maxwell equations have been presented in some new
linear (quasi-Fredholm) integral equation forms. Similar
equations can be derived for all other known integrable
reductions of Einstein equations using the same method without
any its essential modifications.

The alternative representations (\ref{Dressing}) of
the solution $\bigpsi$ of associated linear system in terms of
``in-states'' $\bigpsi_\pm$ and the scattering matrices
$\bchi_\pm$ used here may be considered as some analogue of the
well known dressing methods, developed for solution of various
completely integrable systems (see
\cite{Zakharov-Shabat:1974,Zakharov-Mikhailov:1978,
Belinskii-Zakharov:1978} and the references there). In some points
the present construction can remind also  Krichever's construction
\cite{Krichever:1980} of the ``analogue of d'Alembert formula'' for
the Sine-Gordon equation, as well as a construction of the
homogeneous Hilbert problem and the matrix linear integral equation
form of vacuum Ernst equation presented by Hauser and Ernst
\cite{Hauser-Ernst:1989}. In particular, in \cite{Hauser-Ernst:1989}
the matrices identical to $\bigpsi_\pm$ had been used as
important elements of the developed method.  However, unlike the
mentioned above constructions, closely related with formulations of
various matrix Riemann or Riemann - Hilbert problems, the present
analysis is based on a more detail consideration of some features of
the structure of reduced Einstein equations which allow to reduce
the problem to much more simple, scalar quasi-Fredholm
integral equations.

In comparison with the previously derived (in the framework of
the same, monodromy transform approach) singular integral
equation form of reduced Einstein equations or their
regularizations, which were expressed in terms of conserved
monodromy data \cite{Alekseev:1985,Alekseev:1987}, the new
integral equations are designed mainly for consideration of
initial and boundary value problems for the reduced Einstein
equatons.  Though the ``old'' equations had already provided us
with a principle scheme for solution of the characteristic
initial value problems and the Cauchy problems in the hyperbolic
cases and some boundary problems in the elliptic cases of
integrable reductions of Einstein equations
\cite{Alekseev:1987,Alekseev:1993}, the new integral equations
are obviously better adapted at least for solution of
characteristic initial value problems. The scalar kernels and
the right hand sides of the new equations carry more explicit
information about characteristic initial data, because they are
expressed in terms of ``dynamical'' monodromy data evolving
along the characteristics and being determined completely in
terms of a given characteristic initial data for the fields.

It is interesting to note, that the applicability of the
presented here the ``integral evolution equation'' form of
reduced Einstein equations is restricted from the beginning by
the condition that the analytical (in terms of the coordinates
$\xi$ and $\eta$) local (i.e. near the point where the boundary
characteristics intersect) solutions are considered only.  This
restriction leaves out of our consideration some kinds of
characteristic initial value problems which correspond to physically
interesting enough situations such as, for example, a collision
of plane gravitational waves propagating with distinct
wavefronts on the Minkowski background. In this case, the
regularity of solutions which take place in some globally
defined coordinates is not compatible with their analyticity in
terms of the ``geometrically defined coordinates'' $\xi$ and
$\eta$ near the point of the wavefronts collision
\cite{Griffiths:1991}.  Therefore, to consider a collision
of such waves we need to refine essentially on our methods.
Fortunately the monodromy transform approach, which farther
development was presented in this paper, admits an appropriate
generalization. This generalization was found very recently in
the author's collaborated paper
with J.B.Griffiths \cite{Alekseev-Griffiths:2001} where the
generalized system of linear ``integral evolution equations''
was derived and a method for direct solving of the
characteristic initial value problem for given characteristic
initial data for colliding plane gravitational or gravitational
and electromagnetic waves propagating with distinct wavefronts
on the Minkowski background was presented.

It is necessary to note here also, that in the cases of a
Cauchy problem for hyperbolic reductions or boundary problems
for elliptic reductions of Einstein equations, the interrelations
between the initial or boundary data for the fields and the
functional parameters in the kernels and coefficients of the
integral equations derived in the present paper turn out to be
more complicate than in the characteristic case. Therefore, for
solutions of these problems there is no yet a similar more
simple way than suggested  in \cite{Alekseev:1987,Alekseev:1993}
general scheme for consideration of such problems using the
linear singular integral equations whose construction is based
on the conserved monodormy data.

\section*{Acknowledgments}
The author would like to express his gratitude to Prof.
J.B.~Griffiths for many useful discussions of the colliding
plane wave problems and related topics in General Relativity and
to the Mathematical Department of Loughborough University for
hospitality during his short scientific visits, supported by a
grant from EPSRC, when the present work was begun.  This work
was supported partly by the INTAS (grant 99-1782), RFBR
(the grants 99-01-01150 and 99-02-18415) and by the Russian
Academy of Sciences (the program ``Nonlinear dynamics and
solitons'').

\end{document}